# Cell inactivation by diverse ions along their tracks

Pavel Kundrat[1], Milos Lokajicek, Hana Hromcikova

Institute of Physics, Academy of Sciences of the Czech Republic
Na Slovance 2, CZ-182 21 Praha 8, Czech Republic

**Abstract**
Irradiation of cell monolayers by monoenergetic ions has made it possible to establish survival curves at individual values of linear energy transfer. The two-step model of radiobiological mechanism proposed recently by Judas and Lokajicek (Judas L., Lokajicek M., 2001: Cell inactivation by ionizing particles and the shapes of survival curves. J. Theor. Biol. 210 (1), 15-21., doi:10.1006/jtbi.2001.2283) has then enabled to show that some significant deviations from the generally used linear-quadratic model should exist at higher values of linear energy transfer, which has been also demonstrated experimentally. However, the new model has been expressed in the form being applicable rightfully to low-dose parts of survival curves only. It has been now reformulated to be applicable in analyses of whole survival curves. Inactivation probabilities after different numbers of particles traversing cell nuclei (chromosomal systems) may be then derived from experimental data. Analyses of published data obtained in irradiating by protons and deuterons will be presented and discussed.

**Keywords:** radiobiological mechanism, mathematical modelling, cell inactivation, linear energy transfer

## 1. Introduction

Radiotherapy of tumours increases its potential by using accelerated protons and other ions. However, efficient use of these new irradiation modes requires detailed knowledge of inactivation characteristics at different parts of ion tracks. Physical parameters of beam particles, e.g. their linear energy transfer (LET) values, are changing significantly especially in their Bragg regions. The corresponding inactivation characteristics may be derived from experiments in which cell monolayers are being irradiated by monoenergetic ions.

The ratio of surviving cells in dependence on applied dose $d$ is being expressed generally as

$$s(d) = \exp(-\sigma(d)) . \qquad (1)$$

The function $\sigma(d)$, i.e. the cell survival curve in a semi-logarithmic plot, is then being represented commonly by a simple parabola (the so-called linear-quadratic, LQ model)

$$\sigma(d) = \alpha d + \beta d^2 . \qquad (2)$$

It has been shown by Judas and Lokajicek (2001) that at higher LET values and especially in low-dose regions the survival curves must exhibit significant deviations from the predictions of the standard LQ model, which has been demonstrated also experimentally in an independent way, see Schettino et al. (2001). The corresponding deviations exhibited also the earlier data presented by Belli et al. (1998) and Folkard et al. (1996), which was, however, regarded earlier always as the consequence of statistical experimental errors.

---

[1] Corresponding author.
E-mail address: Pavel.Kundrat@fzu.cz
Postal address: Pavel Kundrat, Institute of Physics, Academy of Sciences of the Czech Republic,
Na Slovance 2, CZ-182 21 Praha 8, Czech Republic
Telephone number: +420 266 052 665, fax number: +420 286 585 443

The formulation of the model in (Judas and Lokajicek 2001), indicating the deviations from the LQ model, has enabled to apply it to the low-dose part of survival curves at rather high LET values only. The contemporary approach makes it possible to apply the model to whole survival curves practically for all not-low LET radiation kinds. In Sec. 2 the basic features of the given two-step model will be described and consequences following for the low-dose parts of survival curves will be shown. Mathematical formulation allowing the model to be applied to whole survival curves will be described in Sec. 3 and the possibilities how to establish the corresponding inactivation characteristics will be discussed. In Sec. 4 the model will be applied to experimental survival curves established in different parts of proton and deuteron track ends, which have been published in the literature. The results will be discussed in Sec. 5. Some concluding remarks will be given in Sec. 6.

## 2. Inactivation of individual cells and the two-step model

If the dose is delivered to cell population in a time shorter than several minutes, as it is usual in radiotherapy, the inactivation mechanism for high-LET radiation may be divided into two subsequent steps. First, a number of energy amounts is transferred to a given cell nucleus (or to a chromosomal system) by individual ionizing particles and corresponding damages of chromosomal DNA are formed.

The cell attempts to repair the damage caused by all the traversing particles. One is entitled to expect that the probability of an unsuccessful repair, i.e. the inactivation probability, depends on total energy transferred to a given cell nucleus. In experiments with cell monolayers irradiated by monoenergetic ions, practically the same energy amount is transferred to cell nucleus during each ion passage; the energy amount being proportional to given LET value. The survival curve can be then expressed in the form (Judas and Lokajicek, 2001)

$$s_\lambda(d) = 1 - \sum_k P_k(\lambda, d) p_k^{(i)}(\lambda) \qquad (3)$$

where $p_k^{(i)}(\lambda)$ are inactivation probabilities after $k$ energy transfers to a cell nucleus, $\lambda$ stands for particle LET value, and $P_k(\lambda,d)$ are probabilities that $k$ energy amounts have been transferred at dose $d$. As the distribution of particles in the beam is quite random one can use the Poisson distribution, i.e.

$$P_k(\lambda, d) = \frac{(hd)^k}{k!} e^{-hd} , \qquad (4)$$

where the average number $h$ of traversing particles per unit dose is given by

$$h = \frac{C\sigma}{\lambda} \qquad (5)$$

and $\sigma$ denotes effective cross section of cell nucleus (or chromosomal system); conversion constant $C = 6.24$ keV/Gy/μm$^{-3}$.

If all functions of $d$ have been expanded in power series (Judas and Lokajicek, 2001) it is possible to write

$$\sigma(d) = \sum_k \alpha_k d^k \qquad (6)$$

where (omitting index $\lambda$ for sake of brevity)

$$\alpha_1 = hp_1^{(i)},$$
$$\alpha_2 = h^2[(1-p_1^{(i)})^2 - (1-p_2^{(i)})]/2,$$
$$\alpha_3 = h^3[-2(1-p_1^{(i)})^3 + 3(1-p_1^{(i)})(1-p_2^{(i)}) - (1-p_3^{(i)})]/6,$$
$$\alpha_4 = h^4[6(1-p_1^{(i)})^4 - 12(1-p_1^{(i)})^2(1-p_2^{(i)}) + 3(1-p_2^{(i)})^2 \qquad (7)$$
$$+ 4(1-p_1^{(i)})(1-p_3^{(i)}) - (1-p_4^{(i)})]/24,$$
$$\alpha_5 = h^5[-24(1-p_1^{(i)})^5 + 60(1-p_1^{(i)})^3(1-p_2^{(i)}) - 30(1-p_1^{(i)})(1-p_2^{(i)})^2$$
$$- 20(1-p_1^{(i)})^2(1-p_3^{(i)}) + 10(1-p_2^{(i)})(1-p_3^{(i)}) + 5(1-p_1^{(i)})(1-p_4^{(i)})$$
$$- (1-p_5^{(i)})]/120,$$

and similarly for higher terms.

If the probabilities were combined in geometrical way, i.e.
$$1 - p_k^{(i)} = (1 - p_1^{(i)})^k, \qquad (8)$$
which would correspond to independent and instantaneous inactivation effects of individual particles, the function $\sigma(d)$ would pass into a straight line. To respect the synergetic or saturation effects of higher particle numbers, which corresponds to accumulation of DNA damages caused by individual particles, one should write in a general case, e.g.,
$$1 - p_k^{(i)} = (1 - p_{k-1}^{(i)})(1 - p_1^{(i)})(1 + \varepsilon_k), \qquad (9)$$
where small parameters $\varepsilon_k$ ($k>1$) may differ from zero. In the simplest case when $\varepsilon_k = \varepsilon$, one can write
$$\alpha_1 = hp_1^{(i)},$$
$$\alpha_2 = -h^2(1-p_1^{(i)})^2 \varepsilon/2, \qquad (10)$$
$$\alpha_3 = h^3(1-p_1^{(i)})^3 (\varepsilon - \varepsilon^2)/6, \ ...$$

The last relations indicate that at least for low value of $d$ the survival curve should differ significantly from a simple parabola described by the LQ model. The signs of $\alpha_k$ terms alternate from $k = 1$ if $\varepsilon > 0$ and from $k = 2$ if $\varepsilon < 0$.

However, the given approach demonstrating directly deviations from the LQ model can be applied to low values of dose only, i.e. to low average number of traversing particles. The modifications ensuring the full applicability of the model also for higher dose values will be described in the next section.

### 3. Non-polynomial application of the two-step model

The polynomial expansion of survival curve cannot be applied practically to higher values of $d$ as the average number of beam particles traversing a cell nucleus becomes too high. A polynomial of very high order would have to be used, which brings numerical problems even if large computers are made use of; for a detailed discussion, see Kundrát (2003).

In such a case it is, however, possible to start with fitting experimental cell survival curves directly to the basic equations (1) and (3) where the parameter $\sigma$ in Eq. (3) may be taken as a free parameter together with all parameters $p_k^{(i)}(\lambda)$ needed for a corresponding dose value; the parameter number being surely higher than $k^{av} = hd$, which is the average number of beam particles hitting a sensitive part of cell nucleus in irradiated cell population; e.g., $k^{av} \approx 30$ for $\sigma = 10\ \mu m^2$, $\lambda = 10$ keV/μm, $d = 5$ Gy.

However, the actual number of free parameters may be significantly diminished if one realizes that the following condition should be fulfilled:

$$0 = p_0^{(i)} \leq p_1^{(i)} \leq \ldots \leq p_k^{(i)} \leq p_{k+1}^{(i)} \leq \ldots \leq 1 . \tag{11}$$

It is then possible to introduce a non-decreasing function $\overline{p}_\lambda(k)$ depending on a significantly lower number of free parameters. In the following analysis we shall write

$$\overline{p}_\lambda(k) = d[1 - \exp(-(ak)^f)] + (1-d)[1 - \exp(-(ck)^e)] \tag{12}$$

where all free parameters $a, \ldots, f$ are non-negative (and $d < 1$); all free parameters depending not only on $\lambda$, but also on particle kind.

### 4. Model analysis of proton and deuteron data

The general model scheme described in the preceding section has been applied to experimental data obtained by Belli et al. (1998) in irradiating Chinese hamster V79 cells by monoenergetic protons. Model calculations of cell survival curves together with experimental data for different values of LET are shown in Figure 1. The $\chi^2$ values depend only weakly on the value of effective cross section $\sigma$ in the interval $10 \,\mu\text{m}^2 \leq \sigma \leq 20 \,\mu\text{m}^2$. The values of other free parameters for individual LET values are given together with corresponding $\chi^2$ values in Table 1; they correspond to the effective cross section $\sigma = 12.8 \,\mu\text{m}^2$ (to be compared with the average geometrical cross section for the V79 nucleus $\sigma = 134 \,\mu\text{m}^2$, as reported by Belli et al. (1998)). The increase of inactivation probabilities with the number of traversing particles, i.e. the $k$-dependences of functions $\overline{p}_\lambda(k)$, for different LET values $\lambda$ are shown in Figure 2.

A similar analysis has been performed for cell survival data obtained by Folkard et al. (1996) for irradiation of V79 cells by low-energy deuterons. The same value of effective cross section $\sigma$ has been assumed, $\sigma = 12.8 \,\mu\text{m}^2$; the values of other free parameters are given in Table 2. The comparison of model results with experimental data is shown in Figure 3. The inactivation probabilities for different LET values are presented in Figure 4.

### 5. Discussion

Cell inactivation probabilities after different numbers of traversing particles are capable to reflect important characteristics of radiobiological mechanism. They represent a solid basis for detailed mathematical modelling of complex processes involved in cell inactivation by radiation.

The parameterization used in our analysis, Eq. (12), consists of a weighted sum of two monotonous functions. It has been necessary to make use of such form in order to reproduce the given experimental data precisely. In order to keep the number of free parameters small, the function $[1-\exp(-(ak)^b)]$ has been used in both terms, with different parameter values. The effective number of free parameters has been reduced further by putting $e = 5.0$ for proton and $e = 2.0$ for deuteron survival curves, respectively.

For both protons and deuterons, the derived inactivation probabilities exhibit rather complex behaviour. They indicate that at least two different kinds of inactivation mechanism have shared in the final effect. The first mechanism seems to be responsible for inactivation after lower numbers of energy transfer events. It may be interpreted as a direct inactivation effect. The other mechanism, manifesting itself from approximately $k = 5$ traversing particles, may be explained as the combination of chromosomal damage with lowered repair capability when the cell has been burdened by greater amount of transferred energy. A kind of saturation effect seems also to be related to the first kind of mechanism, see Figure 2.

The given analysis is to be regarded as preliminary only, as the results have been derived from a rather small set of experimental data having been available.

Nevertheless, the proposed model has enabled a greater insight into the radiobiological mechanism. Some more detailed characteristics might be, of course, revealed if a more sophisticated model for the increase of inactivation probabilities with traversing particle numbers were used and a larger set of systematic experimental data were available.

**6. Conclusion**

The probabilistic two-stage model of radiobiological mechanism enables to describe not only the global shape of cell survival curves but also their detailed behaviour, which might be especially important in case of fractionated irradiation where even small deviations from the global shape are being largely amplified. Moreover, the given model enables one to derive the values of cell inactivation probabilities after different numbers of particles traversing cell nucleus, which opens the way towards detailed realistic modelling of complex physical, chemical and biological processes involved in the radiobiological mechanism. The given results represent also a necessary basis for a more detailed determination of inactivation of cell population irradiated by non-monoenergetic ions, which should contribute to treatment planning in clinical hadron radiotherapy.

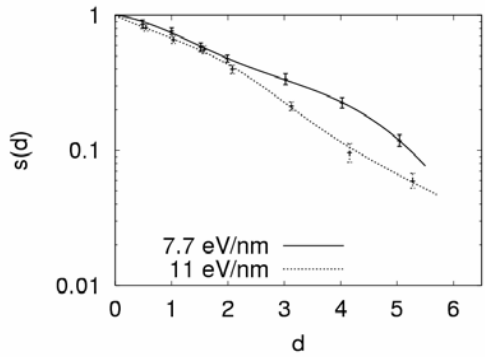
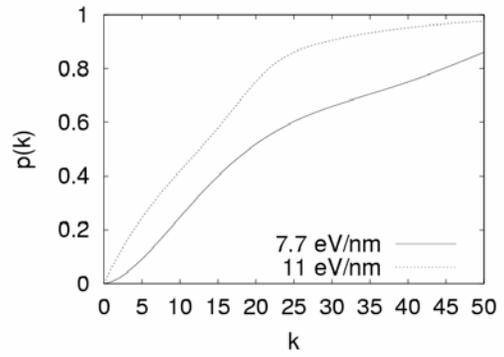
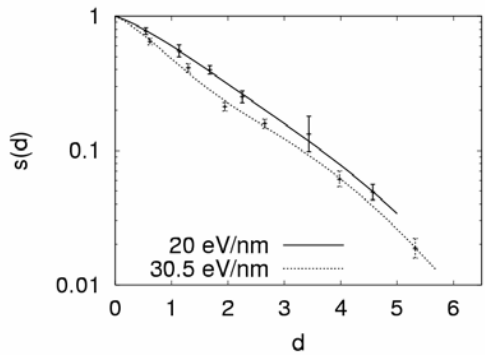
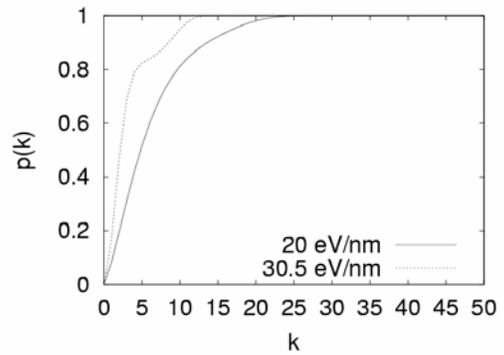
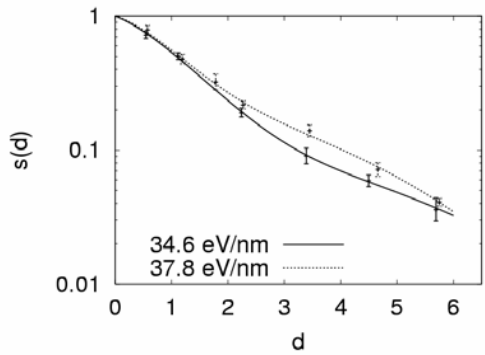
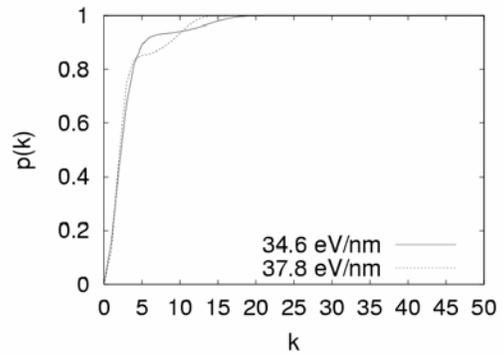

**Figure 1: Survival curves for V79 cells irradiated by low-energy protons. Experimental data established by Belli et al. (1998) – points, fitted by the probabilistic model (lines).**

**Figure 2: Increase of cell inactivation probabilities $p_k^{(i)}$ with number $k$ of protons traversing the chromosomal system.**

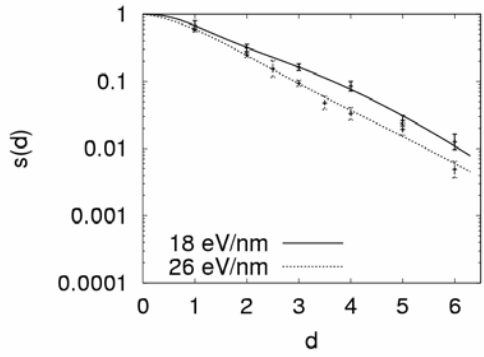
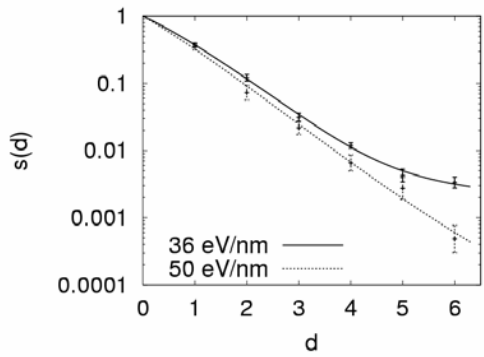
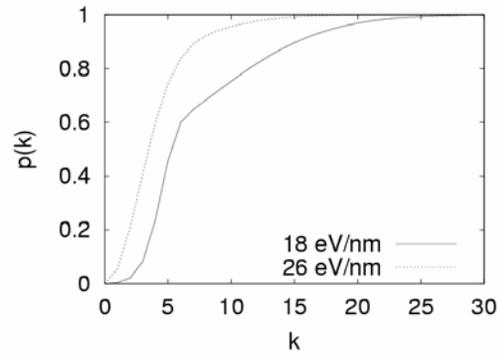
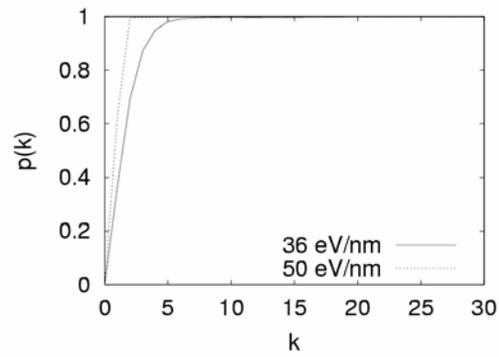

**Figure 3:** Model calculations of survival curves for V79 cells irradiated by low-energy deuterons. Experimental data from Folkard et al. (1996).

**Figure 4:** Increase of cell inactivation probabilities $p_k^{(i)}$ with number $k$ of traversing deuterons.

| $\lambda$ [keV/μm] | 7.7 | 11 | 20 | 30.5 | 34.6 | 37.8 |
|---|---|---|---|---|---|---|
| $d$ | 0.68 | 0.80 | 0.92 | 0.82 | 0.93 | 0.85 |
| $a$ | 0.06 | 0.07 | 0.17 | 0.46 | 0.39 | 0.46 |
| $f$ | 1.66 | 0.99 | 1.35 | 1.88 | 1.75 | 2.36 |
| $c$ | 0.02 | 0.05 | 0.06 | 0.10 | 0.07 | 0.10 |
| $e$ | 5.00 | 5.00 | 5.00 | 5.00 | 5.00 | 5.00 |
| $\chi^2$ | 0.77 | 1.26 | 0.53 | 3.61 | 0.79 | 1.79 |

**Table 1: Values of model parameters describing the increase of inactivation probabilities with number of protons traversing the chromosomal system, Eq. (12), for different LET values. Chromosomal system cross-section $\sigma$ = 12.8 μm$^2$.**

| $\lambda$ [keV/μm] | 18 | 26 | 36 | 50 |
|---|---|---|---|---|
| $d$ | 0.51 | 0.83 | 0.99 | 0.99 |
| $a$ | 0.21 | 0.27 | 0.57 | 1.00 |
| $f$ | 5.00 | 2.06 | 1.38 | 4.97 |
| $c$ | 0.08 | 0.12 | 0.05 | 0.24 |
| $e$ | 2.00 | 2.00 | 2.00 | 2.00 |
| $\chi^2$ | 1.66 | 3.30 | 1.51 | 2.85 |

**Table 2: Values of model parameters describing the increase of inactivation probabilities with number of deuterons traversing the chromosomal system, Eq. (12), for different LET values. Chromosomal system cross-section $\sigma$ = 12.8 μm$^2$.**